\documentclass[aps,floatfix,pra,reprint,superscriptaddress]{revtex4-2}

\usepackage{amsmath}
\usepackage{amssymb}
\usepackage{amsthm}
\usepackage{bbm}
\usepackage{color}
\usepackage{float}
\usepackage[T1]{fontenc}
\usepackage{graphicx}
\usepackage{hyperref}
\usepackage[utf8]{inputenc}
\usepackage{physics}
\usepackage{times}
\usepackage{txfonts}
\usepackage{xcolor}

\hypersetup{
    colorlinks=true,linkcolor=blue,citecolor=blue,
    filecolor=blue,urlcolor=blue,breaklinks=true
}

\newcommand{\orcid}[1]{\href{https://orcid.org/#1}
{\includegraphics[width=7pt]{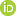}}}

\theoremstyle{definition}
\newtheorem{definition}{Definition}[section]

\def\be{\begin{equation}}
\def\ee{\end{equation}}

\def\bc{\begin{center}}
\def\ec{\end{center}}
\def\bal{\begin{align}}
\def\eal{\end{align}}

\begin{document}

\title{
  Multipartite entanglement sudden death and birth in randomized
  hypergraph states
}

\author{Vinicius Salem\orcid{0000-0002-1768-8783}}
\email{salemfisica@gmail.com}
\affiliation{
  Programa de P\'os-Gradua\c{c}\~{a}o em Ci\^{e}ncias/F\'{i}sica,
  Universidade Estadual de Ponta Grossa,
  84030-900 Ponta Grossa, Paran\'a, Brazil
}

\author{Alison A. Silva\orcid{0000-0003-3552-8780}}
\email{alisonantunessilva@gmail.com}
\affiliation{
  Programa de P\'os-Gradua\c{c}\~{a}o em Ci\^{e}ncias/F\'{i}sica,
  Universidade Estadual de Ponta Grossa,
  84030-900 Ponta Grossa, Paran\'a, Brazil
}

\author{Fabiano M. Andrade\orcid{0000-0001-5383-6168}}
\email{fmandrade@uepg.br}
\affiliation{
  Programa de P\'os-Gradua\c{c}\~{a}o em Ci\^{e}ncias/F\'{i}sica,
  Universidade Estadual de Ponta Grossa,
  84030-900 Ponta Grossa, Paran\'a, Brazil
}
\affiliation{
  Departamento de Matem\'{a}tica e Estat\'{i}stica,
  Universidade Estadual de Ponta Grossa,
  84030-900 Ponta Grossa, Paran\'a, Brazil
}

\date{\today}

\begin{abstract}
We introduce and analyze the entanglement properties of randomized
hypergraph states, as an extended notion of the randomization
procedure in the quantum logic gates for the usual graph states,
recently proposed in the literature.
The probabilities of applying imperfect generalized controlled-$Z$
gates simulate the noisy operations over the qubits.
We obtain entanglement measures as negativity, concurrence, and genuine
multiparticle negativity, and show that  entanglement exhibits a
non-monotonic behavior in terms of the randomness parameters, which is a
consequence of the non-uniformity of the associated hypergraphs,
reinforcing the claim that the entanglement of randomized graph states is
monotonic since they are related to $2$-uniform hypergraphs.
Moreover, we observed the phenomena of entanglement sudden death
and entanglement sudden birth in RH states.
This work reveals a connection between the non-uniformity of
hypergraphs and loss of entanglement.\\

\noindent DOI:
\href{https://doi.org/10.1103/PhysRevA.109.012416}
{10.1103/PhysRevA.109.012416}

\end{abstract}

\maketitle{}

\section{Introduction}
Multiparticle entanglement is a resource in quantum information
processing (QIP), providing advantages over its classical counterparts,
and is relevant in many physical applications
\cite{RMP.81.865.2009,RMP.80.517.2008}.
Hypergraph states (HS) concern a family of multiparticle quantum states
that are genuinely maximally entangled
\cite{JPA.48.095301.2015,JPA.47.415304.2014,JPA.47.335303.2014,
PRA.87.022311.2013,NJP.15.113022.2013,PRA.79.052304.2009} and have been
the subject of intense research
\cite{JPA.56.375302.2023,arXiv:2301.11341,PRA.106.012410.2022}.
This family generalizes the well-known family of graph states (GS),
which are $2$-uniform HS \cite{JPA.51.045302.2017}, that rely on the
center of interest in quantum information protocols
\cite{PRApp.12.054047.2019}, show potential for foundational studies
\cite{JPA.50.245303.2017}, and violate local realism
\cite{JPA.51.125302.2018,PRL.116.070401.2016}.

Controlled systems that demand quantum measurements, such as quantum
logic gates, still present some unavoidable degree of noise and the
application of the quantum gates not always can be performed with
success.
While $1$-qubit gates can be built with fidelities higher than
$99\%$, $2$-qubit entangling gates hardly reach $93\%$, and this becomes
worse for $3$-qubit gates, and so on
\cite{PRX.11.021058.2021,PRL.131.210603.2023,PRA.83.042314.2011,
PRL.105.230503.2010}.
So, an understanding of random quantum states and their
generation in the presence of noise are of current interest
\cite{PRR.2.023306.2020}.
Quantum decoherence resulting from low-fidelity gates, environmental
factors, and inter qubit interactions \cite{Inproceedings.2000.Joos}
represents a significant challenge in quantum computing.
Indeed, one of the main goals of QIP is to assure a way to perform
universal quantum computation in the most feasibly robust way
\cite{PRA.86.050301.2012}.
A demonstration of the deleterious effects of decoherence on entangled
states is the entanglement sudden death (ESD), which refers to the
complete loss of entanglement in a finite time, regardless of the
asymptotic nature of the loss of system coherence
\cite{S.323.598.2009,PRL.93.140404.2004}.
The phenomenon of ESD has been observed experimentally in different
physical systems
\cite{S.316.579.2007,PRL.99.180504.2007,PRL.104.100502.2010,
PRL.109.150403.2012,NP.8.117.2011}, and is something we want to avoid
for any practical purpose of using entangled states.

Most of the studies investigating the ESD phenomenon concern bipartite
systems.
For multiparticle systems, we must take into account the concept of
genuine multiparticle entanglement (GME)  \cite{PRA.62.062314.2000}.
In this paper, we discuss noisy multipartite gates by introducing the
concept of randomized hypergraph (RH) states as a generalization of the
concept of randomized graph (RG) states \cite{PRA.89.052335.2014}.
We show that this randomization procedure of nonuniform hypergraphs
lead to RH states that break the monotonic behavior of bipartite
entanglement (BE) and GME as a function of the randomness
parameters, reinforcing the claim raised in
Ref. \cite{PRA.89.052335.2014} that the entanglement of RG states is
monotonic, since these latter states are 2-uniform HS.
Moreover, we also show the manifestation of ESD, as well as entanglement
sudden birth (ESB) \cite{PRL.101.080503.2008}, in RH states as a
function of the randomness parameters.
We note that these cases occur for non-uniform hypergraphs, the
ones where hyperedges of different cardinalities are present.

\section{Hypergraph states}
\begin{figure}[t]
  \centering
  \includegraphics[width=\columnwidth]{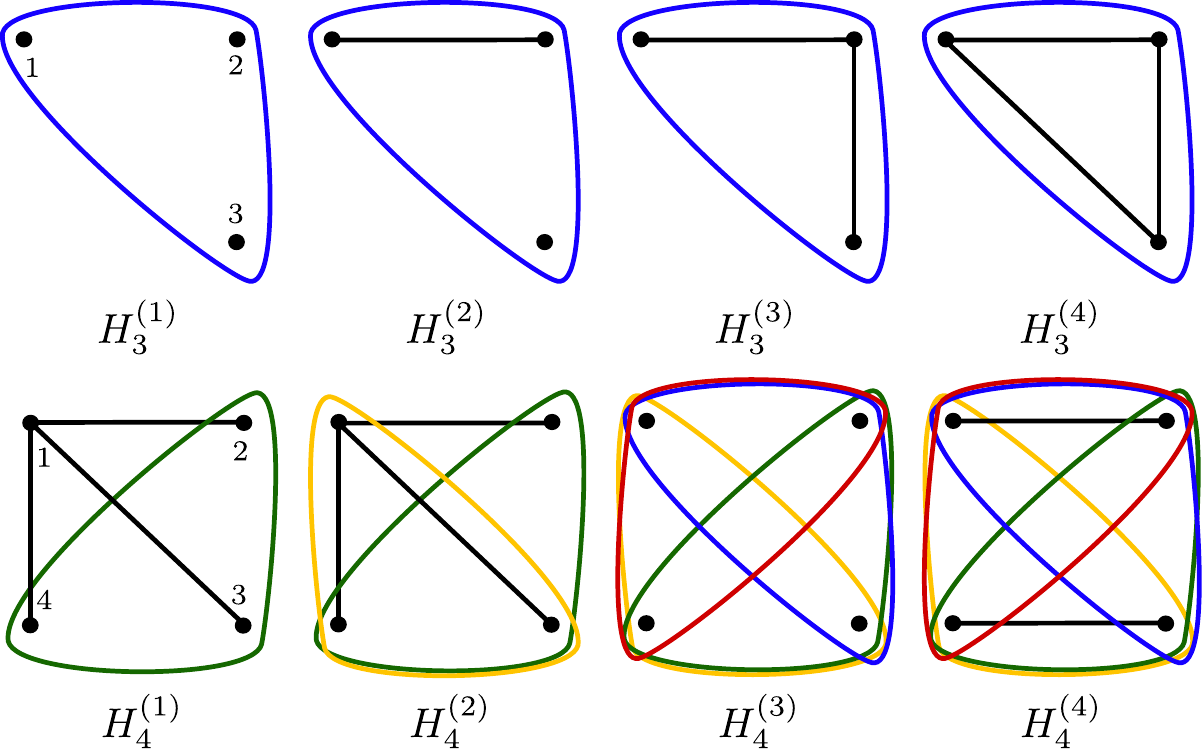}
  \caption{
    Hypergraphs that are of interest in this paper.
    The hypergraphs $H_4^{(1)}$, $H_4^{(2)}$, and $H_{4}^{(3)}$ are cases
    of special interest because their reduced single-qubit matrices are
    maximally mixed \cite{JPA.47.335303.2014}.
    $H_{n}^{(i)}$ represents the $i$th hypergraph on $n$ vertices.
  }
  \label{fig:fig1}
\end{figure}

A hypergraph $H=(V,E)$ is defined as a pair consisting of a finite set
of vertices $V=\{v_1,v_2,\ldots,v_n\}$ and a set of hyperedges
$E \subset 2^{V}$, with $2^{V}$ the power set of $V$
\cite{Book.2001.Berge}.
A hyperedge that encloses $k$ vertices has cardinality $k$ and if a
hypergraph has all hyperedges with cardinality $k$, the hypergraph is
$k$-uniform.
An ordinary graph is a $2$-uniform hypergraph.
If a hypergraph $F$ is obtained from $H$ by deleting hyperedges, in such
a way that $V_F=V_H$, then $F$ is called a subhypergraph of $H$
\cite{Book.Lovasz.1986}.
In this latter case, we say that $F$ spans $H$.
Examples of hypergraphs are shown in Fig. \ref{fig:fig1}.
Given a hypergraph $H$ on $n$ vertices, the correspondent HS, denoted by
$\ket{H}$, is obtained by assigning a qubit for every vertex of the
hypergraph in the state $\ket{+}=(\ket{0}+ \ket{1})/\sqrt{2}$, in such a
way that the initial state is $\ket{+}^{\otimes n}$, followed by the
application of a non-local multiqubit gate $C_e$ acting on the Hilbert
spaces associated with vertices $v_i\in e$, which is a
$2^{|e|} \times 2^{|e|}$ diagonal matrix given by
$C_e= \mathbbm{1} -2\ket{1 \ldots 1}\bra{1 \ldots 1}$,
where $e \in E$ represents a hyperedge, $|e|$ is the cardinality of the
hyperedge $e$, and $\mathbbm{1}$ is the identity matrix.
Thus, a HS $\ket{H}$ is a pure quantum state defined as
\cite{NJP.15.113022.2013}
\begin{equation}
  \ket{H} = \prod_{e \in E} C_{e} \ket{+}^{\otimes n}.
\end{equation}

\section{Randomized Hypergraph States}
Instead of applying the perfect usual $C_e$ gate to obtain a pure HS, we
define a randomization operator $R_P$ which introduces the application
of probabilistic gates $\Lambda_{p_{|e|}}^{e}$ to the state
$\ket{+}^{\otimes n}$, where $P=\{p_k\}_{k=1}^{n}$ is a set of
probabilities (or randomness parameters).
Thus, we consider a noise implementation of the generalized $C_e$ gate,
in which with probability $p_{|e|}$ the gate succeeds (a hyperedge is
created among the qubits) and with probability $(1-p_{|e|})$ the gate
fails, which has the same effect as an identity operator
\cite{PRA.89.052335.2014}.
This allowed us to simulate the presence of noise in the construction of
the states that can be used for quantum computational universality for
hypergraphs \cite{SR.9.1.2019}.
All the results are obtained concerning the randomness parameter
$p_{|e|}$, which can be implicitly linked with time and is deliberately
left indeterminate to accommodate various possible noise behaviors.
As an example, $p_{|e|}=e^{-\kappa_{|e|} t}$, where $t$ might
represent the time duration over which a quantum gate is actively
applied or attempted and $\kappa_{|e|}$ is the decay constant.
The decay could then signify a reduction in the probability of the
success of gate as time progresses.

\begin{figure}[t]
  \centering
  \includegraphics[width=\columnwidth]{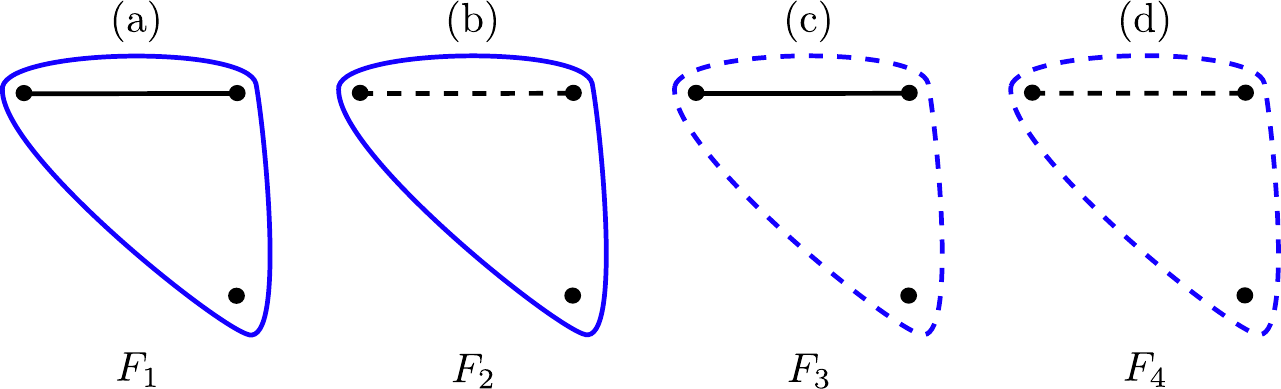}
  \caption{
      Process of randomization of the $\ket{H_{3}^{(2)}}$.
      Dashed lines represent noisy $C_e$ gates, in which with probability
      $p_{|e|}$ the gate succeeds and with probability $(1-p_{|e|})$
      the gate fails.
      When the gate fails, it has the same effect as an identity operator.
      Continuum lines are a successful creation of hyperedges.
      (a) $F_1$ subhypergraph with probability $p_2p_3$;
      (b) $F_2$ subhypergraph with probability $(1-p_2)p_3$;
      (c) $F_3$ subhypergraph with probability $p_2(1-p_3)$;
      and
      (d) $F_4$ subhypergraph with probability $(1-p_2)(1-p_3)$ [see
      Eq. \eqref{eq:Rp}].
    }
  \label{fig:fig2}
\end{figure}

To exemplify the procedure, consider the hypergraph $H_{3}^{(2)}$ in
Fig. \ref{fig:fig1}.
The randomization procedure produces the following state,
\begin{align}
  \label{eq:Rp}
  R_{P}\left(\ket{H_{3}^{(2)}}\right)
  = {}
  &
  \Lambda^{\{1,2\}}_{p_2} \circ \Lambda^{\{1,2,3\}}_{p_3}(\ketbra{+++})
  \nonumber \\
  = {}
    &
      p_2p_3\ketbra{F_{1}}{F_{1}}+(1-p_2)p_3\ketbra{F_2}{F_2}
      \nonumber \\
    &
      +p_2(1-p_3)\ketbra{F_3}{F_3}
      \nonumber \\
    &
      +(1-p_2)(1-p_3)\ketbra{F_4}{F_4},
\end{align}
with $p_2$ and $p_3$ the randomness parameters for the application of
the hyperedges of cardinalities $2$ and $3$, respectively.
The spanning subhypergraphs $F_i$, $i=1,\ldots,4$, are schematically
shown in Fig. \ref{fig:fig2}.
Since all the gates $\Lambda_{p_{|e|}}^{e}$ commute, the order of
application of these gates does not need to be specified.
Due to the application of this randomization operator, we end up having
randomized mixed states, contrary to the pure states that
HS usually represent.
These RH states will be denoted as $\rho^P_H$.
Now we give a formal definition of RH states.
\begin{definition}[Randomized hypergraph state]
  Let $\ket{H}$ be a hypergraph state.
  Its randomization operator is defined as
\begin{equation}
  R_P(\ket{H})=\sum_{F \text{ spans } H}
    \prod_{p_k\in P}
    p_k^{|E_{k,F}|}(1-p_k)^{|E_{k,H} \setminus E_{k,F}|}
    \ketbra{F}{F},
\end{equation}
where $F$ are the spanning subhypergraphs of $H$, $E_{k,H}$ and
$E_{k,F}$ are the sets of hyperedges of $H$ and $F$, respectively, and
$P=\{p_k\}_{k=1}^{n}$ is the set of randomness parameters for hyperedges
of cardinality $k$.
The resulting hypergraph state $\rho_{H}^{P}:= R_P(\ket{H})$ is the
randomized version of $\ket{H}$.
\end{definition}

Thus applying the above definition, we can map an initially pure
hypergraph state $\ket{H}$ onto a mixture of all its spanning
subhypergraph states with probability $p_{k}$, $k=1,\ldots,n$,
making the role of a parameter controlling the amount of entanglement of
a RH state that connects the two extremes cases, for $p_{k}=0$,
an empty hypergraph, and $p_k=1$ for a pure HS.
It is important to emphasize that the randomization is related only to
the hyperedges of the hypergraph initially given.

\begin{figure*}[t]
  \centering
  \includegraphics[width=\textwidth]{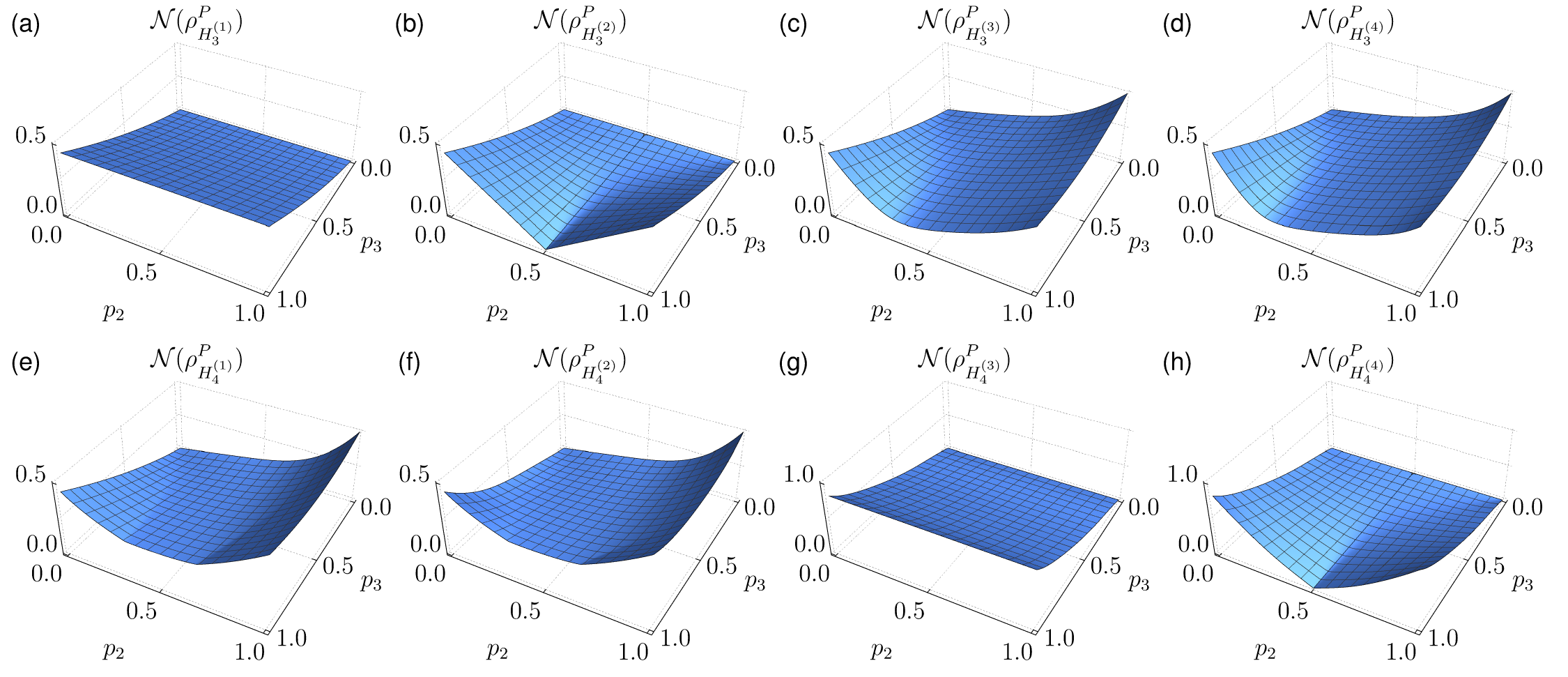}
  \caption{
    Negativity for RH states listed in Fig. \ref{fig:fig1}.
    The notation ``$\{v_1,\ldots\}|\{u_1,\ldots\}$'' represents the
    bipartition used to calculate the negativity.
    (a)--(d) Negativity of 3-qubit RH states for bipartition $\{3\}|\{1,2\}$.
    (e)--(f) Negativity of 4-qubit RH states for bipartition $\{2\}|\{1,3,4\}$.
    (g)--(h) Negativity of 4-qubit RH states for bipartition $\{1,2\}|\{3,4\}$.
  }
  \label{fig:fig3}
\end{figure*}

Physically, it is crucial to know the equivalent classes between states,
since those states share the same entanglement properties.
Two $n$-qubit HS $\ket{H}$ and $\ket{H'}$ are said to be
local unitary (LU) equivalent, and so belonging to the same entanglement
class, if and only if they can be connected by local unitaries
$U_{1},\ldots,U_{n}$, in such a way that
$\ket{H}=U_1 \otimes\ldots\otimes U_n \ket{H'}$.
This property turns out to be quite important, since if we identify
LU equivalent classes of HS, it is possible to know if
they are also equivalent under stochastic local operations and classical
communication \cite{JPA.47.335303.2014}.
A deeper analysis of the equivalent classes of HS up to
six qubits can be found in Ref. \cite{JPA.47.335303.2014}.
As in the case of RG states generated from two LU equivalent GS, RH
states generated from two LU equivalent HS are, in general, not LU
equivalent.
This stems from the fact that the randomization process produces mixed
states with different ranks and therefore they can no longer be
transformed using LU operations.
This fact is an attractive property of RH states, since the
purpose of the randomization is to precisely simulate the noise degree
during the experimental construction of these states in a laboratory.
An interesting explanation concerning the experimental implementation
using hypergraph states can be found in Ref. \cite{PRA.101.033816.2020}.

\section{Bipartite Entanglement}
To illustrate our findings concerning the entanglement properties of RH
states, we have chosen hypergraphs shown in Fig. \ref{fig:fig1}.
Since RH states are mixed states, we use the  negativity
\cite{PRA.65.032314.2002} and the concurrence
\cite{PRL.78.5022.1997} to quantify the amount of BE.
The negativity of a bipartite state $\rho$ is defined as
$\mathcal{N}(\rho)=(\|\rho^{\Gamma_A}\|-1)/2$, where $\rho^{\Gamma_{A}}$
is the partial transpose of $\rho$ with respect to subsystem $A$, and
$\|X\|=\sqrt{X^{\dagger}X}$ is the trace norm.
The concurrence for a mixed bipartite state $\rho$ is defined as
$\mathcal{C}(\rho)=\max\{0,\lambda_1-\lambda_2-\lambda_3-\lambda_4\}$,
where $\lambda_i$ are the eigenvalues, in decreasing order, of the
non-Hermitian matrix $\rho \tilde{\rho}$, with
$\tilde{\rho}=
(\sigma_{y}\otimes \sigma_y) \rho^{*}
(\sigma_y \otimes \sigma_y)$.

\begin{figure*}[t]
  \centering
  \includegraphics[width=\textwidth]{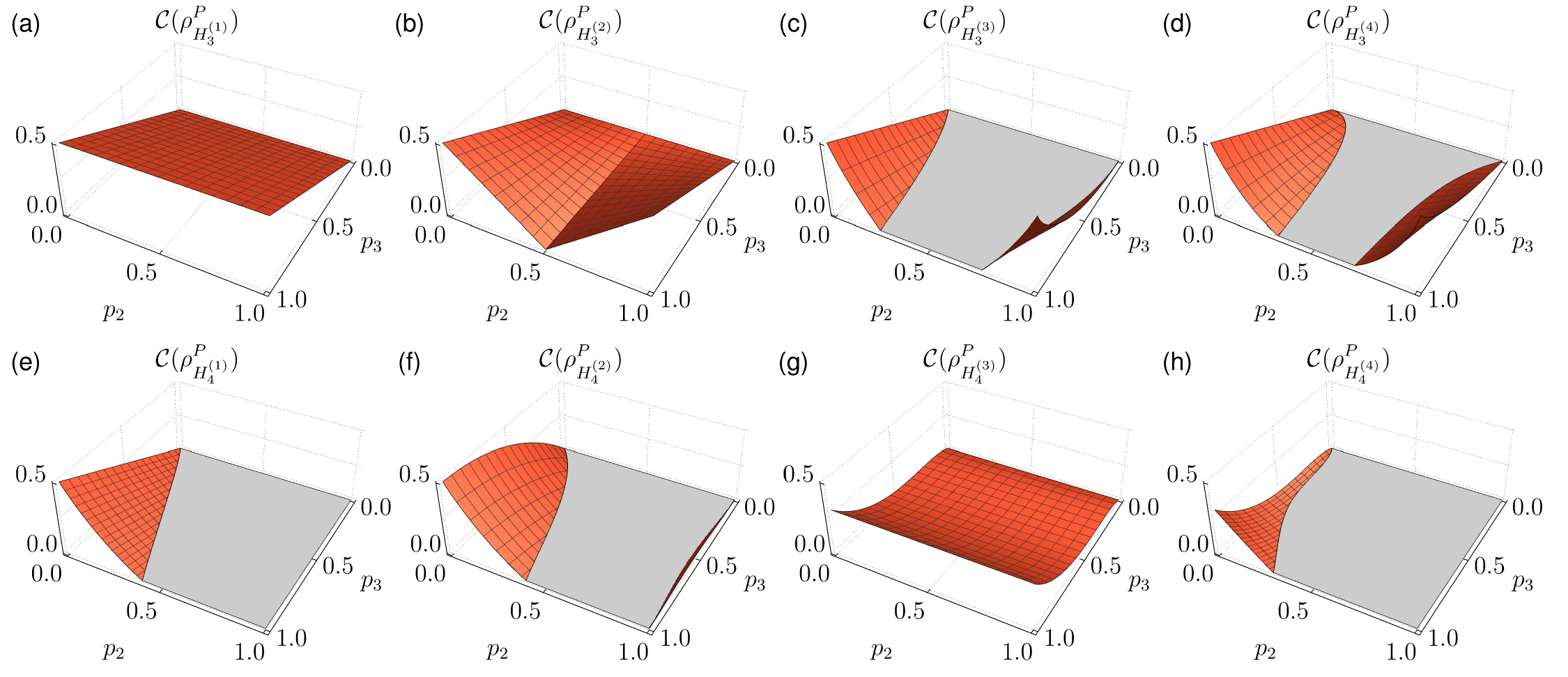}
  \caption{
    Concurrence of RH states listed in Fig. \ref{fig:fig1}.
    (a)--(d) Concurrence of 3-qubit RH states between qubits $1$ and
    $3$;
    (e), (f) concurrence of 4-qubit RH states between qubits $3$ and $4$;
    and
    (g), (h) concurrence of 4-qubit RH states between qubits $1$ and $3$.
    The light gray part indicates the zero value of concurrence.
  }
  \label{fig:fig4}
\end{figure*}

\begin{figure*}[t]
  \centering
  \includegraphics[width=\textwidth]{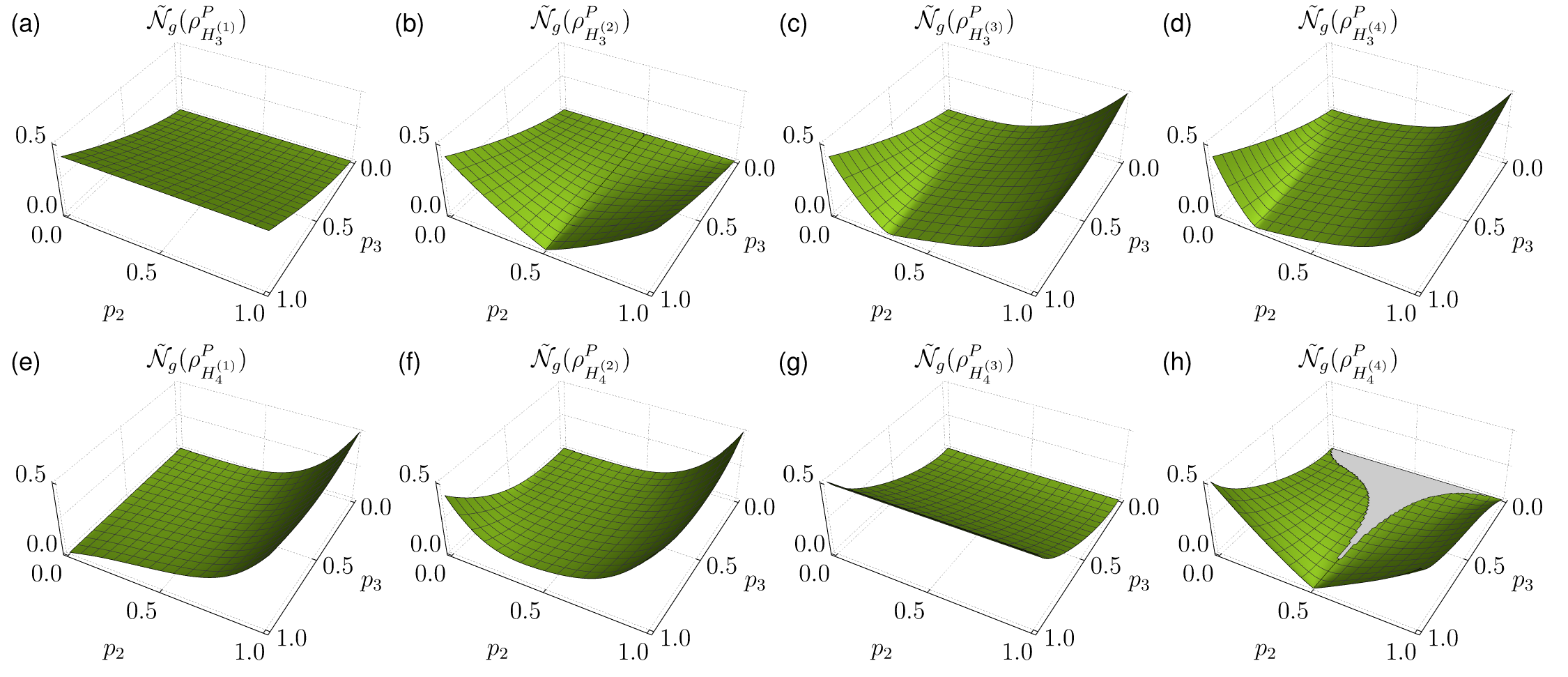}
  \caption{
    GMN of RH states listed in Fig. \ref{fig:fig1}.
    (a)--(d) GMN for 3-qubit RH states.
    (e)--(h) GMN for 4-qubit RH states.
    The light gray part indicates the zero value of GMN.
  }
  \label{fig:fig5}
\end{figure*}

\begin{table}[b]
  \caption{
    Values of $p_2$ for which the ESD and ESB occur for $3$- and
    $4$-qubit RH states listed in Fig. \ref{fig:fig1} for $p_3=1$.
  }
\begin{tabular}{l c c c c c c}
\hline
\hline
RH state              &  ESD    &  ESB    & \hspace{2cm} &
RH state              &  ESD    &  ESB
\\
\hline
$\rho_{H_{3}^{(1)}}^{P}$ &  --     &  --     &             &
$\rho_{H_{4}^{(1)}}^{P}$ & $0.397$ & --
\\
$\rho_{H_{3}^{(2)}}^{P}$ & $0.500$ & $0.500$ &             &
$\rho_{H_{4}^{(2)}}^{P}$ & $0.397$ & --
\\
$\rho_{H_{3}^{(3)}}^{P}$ & $0.239$ & $0.761$ &             &
$\rho_{H_{4}^{(3)}}^{P}$ & --      & --
\\
$\rho_{H_{3}^{(4)}}^{P}$ & $0.315$ & $0.707$ &             &
$\rho_{H_{4}^{(4)}}^{P}$ & $0.288$ & --
\\
\hline \hline
\end{tabular}
\label{tab:tab1}
\end{table}

In Fig.  \ref{fig:fig3} we show the results for the negativity
for the RH states in Fig. \ref{fig:fig1} as a function of the
randomness parameters $p_2$ and $p_3$.
We can observe that the negativity shows a nonmonotonic behavior for
all cases but $\rho_{H_{3}^{(1)}}^{P}$ and $\rho_{H_{4}^{(3)}}^{P}$
[see Figs. \ref{fig:fig3}(a) and  \ref{fig:fig3}(g), respectively],
which show a monotonic behavior.
This is an interesting result because in Ref. \cite{PRA.89.052335.2014}
an observation was made of a monotonic behavior of negativity in terms
of the randomness parameter for all the RG states studied there.
Moreover, the question was raised regarding whether the monotonic
behavior of the entanglement is a common feature for all RG states.
This monotonic behavior of negativity was also observed in
Ref. \cite{LP.31.035201.2021} for multiqubit randomized entangled states
obtained by using random Ising-type entangling operators.
However, our results show that the answer to this question is in the
negative for RH states.
This behavior can be understood by noting that the monotonic behavior
of the negativity stems from the fact that $\rho_{H_{3}^{(1)}}^{P}$
and $\rho_{H_{4}^{(3)}}^{P}$ are states associated with $3$-uniform
hypergraphs, in which there is no mixture of hyperedges of different
cardinalities.
We can say there is a sort of competition for entanglement when applying
the entangling gates of different cardinalities.
Thus, the break of the monotonic behavior of the BE in RH states is a
feature present in those states that are associated with nonuniform
hypergraphs and such behavior could not be observed in RG states, since
any graph is a 2-uniform hypergraph.

In Fig. \ref{fig:fig4}, we present the results for the concurrence for
the RH states in Fig. \ref{fig:fig1}.
As the negativity, the concurrence also manifests a nonmonotonic
behavior of the BE for non-uniform HS.
More interesting is the presence of ESD and ESB, as we can see large
regions where the concurrence vanishes.
In Table \ref{tab:tab1}, we show the values of $p_2$ for the
manifestation of ESD and ESB for the RH states for the case $p_3=1$.
For instance, consider Fig. \ref{fig:fig4}(c) where we can observe the
manifestation of ESD at $p_2=0.239$ and ESB at $p_2=0.761$.
So, again we observe a sort of competition for the entanglement because
of the application of entangling gates of different cardinalities, but
now with impressive results.
We emphasize that the concurrence is the entanglement measure that shows
the most anomalous behavior for the entanglement results.
It is somehow expected since concurrence is calculated using two-qubit
reduced density matrices.

\section{Genuine Multiparticle Entanglement}
Since concurrence and negativity are not able to quantify the amount of
GME in RH states, we employ the concept of genuine multiparticle
negativity (GMN) \cite{PRL.106.190502.2011,JPA.47.155301.2014}, to
quantify the amount of GME.
It is well known that HS are GME states \cite{JPA.47.335303.2014}, which
are quite useful for quantum information protocols \cite{NJP.19.093012.2017,IJQI.04.415.2006,PRA.59.1829.1999}.
The GMN is a versatile tool to characterize multiparticle entanglement,
where its implementation is based on fully decomposed witnesses
$\mathcal{W}$, and it can be directly computed using semidefinite
programming.
One defines the GMN $\tilde{\mathcal{N}}_g(\rho)$ by means of the
optimization problem
\begin{equation}
  \tilde{\mathcal{N}}_g(\rho) = -\min \tr(\rho \mathcal{W}),
\end{equation}
subject to $\mathcal{W}=P_m + Q_m^{T_m}$, with
$0 \leqslant P_m \leqslant \mathbbm{1}$ and
$0 \leqslant Q_m \leqslant \mathbbm{1}$, for all the partitions
$\{m\}|\{\tilde{m}\}$, where $\{\tilde{m}\}$ is the complement of the
part $\{m\}$, and $T_m$ is the partial transpose with respect to $m$.
The results for the GMN of RH states were obtained with the help of the
online program PPTMIXER \cite{PPTMixer}, and are shown in
Fig. \ref{fig:fig5}.
As was the case for BE, the GMN for the RH states of
Fig. \ref{fig:fig1} also shows a nonmonotonic behavior for all cases
except for $\rho_{H_3^{(1)}}^{P}$ and $\rho_{H_4^{(3)}}^{P}$, which are
related to uniform hypergraphs.
It is interesting to comment that the RH state $\rho_{H_3^{(2)}}^{P}$
manifests bipartite and genuine multiparticle ESD for $p_2=0.5$ for all
values of $p_3$.
The RH state $\rho_{H_4^{(4)}}^{P}$ seems to be the most fragile one
as the GMN (and concurrence) is the most affected one by the variation
of the randomness parameters, as it shows a large region of null
entanglement [see Figs. \ref{fig:fig4}(h) and \ref{fig:fig5}(h)].

\section{Conclusion}
To summarize, we have introduced the concept of RH states.
These randomized mixed states have a nonmonotonic behavior of BE
and GME in terms of the randomness parameters, a feature not observed in
RG states \cite{PRA.89.052335.2014}.
The nonmonotonic behavior of entanglement occurs in RH states
associated with non-uniform hypergraphs, while the monotonic behavior is
observed only in uniform RH states, reinforcing the claim that the
entanglement in all RG states is generally monotonic in terms of the
randomness parameter.
Moreover, to support this conclusion we computed the entanglement
properties for all the $27$ equivalence classes under LU transformations
for HS of four qubits \cite{JPA.47.335303.2014} and it was observed the
break of monotonicity of entanglement for non-uniform hypergraphs.
Moreover, the interesting phenomena of bipartite and multiparticle ESD
and ESB for some RH states were observed.
Such behavior is as strong as the difference between the cardinalities
of the hyperedges.
These results and related issues will be reported in a future work \cite{inpreparation}.

This work reveals a connection between the non-uniformity of
hypergraphs and loss of entanglement,
changing the way we can view ESD in multiparticle states.
Observing the results for BE and GME, we can observe that the ESD for
GME can be less frequent, a counter-intuitive result that was recently
observed in Ref. \cite{PRR.5.l032015.2023}.
A precise description of ESD for multiparticle systems remains a
challenging topic and there is still sparse literature.
Such an intricate interplay between noise-induced perturbations and the
capacity for entanglement regeneration emphasizes the resilience of the
system under decoherence, potentially contributing to the development of
more robust and fault-tolerant QIP techniques.

\section*{Acknowledgments}
We thank Danilo Cius for enlightening discussions and the anonymous
Referees for their valuable comments and suggestions.
This work was partially done at ICFO (Institut de Ciencies Fotoniques)
of Barcelona (ES) and was supported by the
Brazilian agencies Conselho Nacional de Desenvolvimento Cient\'ifico e
Te\-cnol\'ogico (CNPq) and
Instituto Nacional de Ci\^{e}ncia e Tecnologia de Informa\c{c}\~{a}o
Qu\^{a}ntica (CNPq, INCT-IQ 465469/2014-0).
It was also financed by the Co\-or\-dena\c{c}\~{a}o de
Aperfei\c{c}oamento de Pessoal de N\'{i}vel Superior (CAPES, Finance
Code 001).
F.M.A. acknowledges CNPq Grant No. 314594/2020-5.

%

\end{document}